\documentclass[aps,prd,a4paper,twocolumn,amsmath,showpacs,superscriptaddress,nofootinbib,preprintnumbers]{revtex4-1}

\usepackage{verbatim}
\usepackage[T1]{fontenc}
\usepackage[utf8]{inputenc}
\usepackage[american]{babel}
\usepackage{epsfig}
\usepackage{graphicx}
\usepackage{booktabs}
\usepackage{multirow}
\usepackage{dcolumn}
\usepackage{amsmath}
\usepackage{mathtools}
\usepackage{amsfonts}
\usepackage{amssymb}
\usepackage{epstopdf}
\usepackage{bm}
\usepackage{siunitx}
\usepackage{braket}
\usepackage{enumitem}
\usepackage{soul}
\usepackage[table]{xcolor}
\usepackage{color}
\usepackage{transparent}
\usepackage{pifont}

\definecolor{navyblue}{rgb}{0.0, 0.0, 0.5}
\definecolor{royalblue}{rgb}{0.25, 0.41, 0.88}
\definecolor{cadmiumgreen}{rgb}{0.0, 0.42, 0.24}
\definecolor{blue-violet}{rgb}{0.54, 0.17, 0.89}
\definecolor{darkviolet}{rgb}{0.58, 0.0, 0.83}
\definecolor{orange(colorwheel)}{rgb}{1.0, 0.5, 0.0}

\usepackage{hyperref}
\hypersetup{
    colorlinks=true, 
    linkcolor=royalblue, 
    citecolor=magenta}

\newcommand\ee{\end{equation}}
\newcommand\be{\begin{equation}}
\newcommand\eea{\end{eqnarray}}
\newcommand\bea{\begin{eqnarray}}

\newcommand{\ns}{n_{\rm s}}
\newcommand{\nrun}{n_{\rm run}}

\usepackage{booktabs}
\usepackage{multirow}
\usepackage{dcolumn}
\usepackage{colortbl}

\definecolor{magenta(process)}{rgb}{1.0, 0.0, 0.56}

\definecolor{darkspringgreen}{rgb}{0.09, 0.45, 0.27}

\definecolor{royalblue(web)}{rgb}{0.25, 0.41, 0.88}

\begin{document}

\title{The impact of theoretical assumptions in the determination of the neutrino effective number from future CMB measurements}  

\author{Ludovico Capparelli}
\email{ludovico.mark.capparelli@roma1.infn.it}
\affiliation{Physics Department and INFN, Universit\`a di Roma ``La Sapienza'', Ple Aldo Moro 2, 00185, Rome, Italy} 

\author{Eleonora Di Valentino}
\email{eleonora.divalentino@manchester.ac.uk}
\affiliation{Jodrell Bank Center for Astrophysics, School of Physics and Astronomy, University of Manchester, Oxford Road, Manchester, M13 9PL, UK}

\author{Alessandro Melchiorri}
\email{alessandro.melchiorri@roma1.infn.it}
\affiliation{Physics Department and INFN, Universit\`a di Roma ``La Sapienza'', Ple Aldo Moro 2, 00185, Rome, Italy} 

\author{Jens Chluba}
\email{jens.chluba@manchester.ac.uk}
\affiliation{Jodrell Bank Center for Astrophysics, School of Physics and Astronomy, University of Manchester, Oxford Road, Manchester, M13 9PL, UK}

\date{\today}

\preprint{}
\begin{abstract}
One of the major goals of future Cosmic Microwave Background measurements is the accurate determination of the effective number of neutrinos $N_{\rm eff}$. Reaching an experimental sensitivity of $\Delta N_{\rm eff} = 0.013$ could indeed falsify the presence of any non-standard relativistic particles at $95 \%$ c.l.. In this paper, we test how this future constraint can be affected by the removal of two common assumptions: a negligible running of the inflationary spectral index $n_{\rm run}$ and a precise determination of the neutron lifetime $\tau_n$. We first show that the constraints on $N_{\rm eff}$ could be significantly biased by the unaccounted presence of a running of the spectral index. Considering the Stage-IV experiment, a negative running of ${\rm d}n/{\rm d}\ln k= - 0.002$ could mimic a positive variation of $\Delta N_{\rm eff} = 0.03$. Moreover, given the current discrepancies between experimental measurements of the neutron lifetime $\tau_n$, we show that the assumption of a conservative error of $\Delta\tau_n \sim 10$s could bring to a systematic error of $\Delta N_{\rm eff} = 0.02$. Complementary cosmological constraints on the running of the spectral index and a solution to the neutron lifetime discrepancy are therefore needed for an accurate and reliable future CMB bound of $N_{\rm eff}$ at percent level.
\end{abstract}

\maketitle

\section{Introduction}

In the past decade, measurements of the Cosmic Microwave Background anisotropies made from satellite experiments as the Wilkinson Microwave Anisotropy Probe (WMAP) \cite{wmap} and Planck \cite{planck2015} have spectacularly confirmed the theoretical expectations of the standard model of structure formation based on inflation, cold dark matter and a cosmological constant.
This nearly perfect agreement between theory and observations is now letting cosmologists to use the CMB data to constrain several aspect of fundamental physics (see e.g. \cite{planck2015}).

Among these parameters, a key observable is the effective neutrino number, $N_{\rm eff}$, that determines the number of relativistic degrees of freedom during the epoch of CMB anisotropies formation, at the last scattering surface (see e.g.~\cite{archi}). A change in the neutrino effective number, affects the epoch of equality and modifies the CMB horizon and damping scales, yielding a characteristic imprint on the CMB~\cite{neff}. The latest measurement made by the Planck satellite \cite{planck2015,plancknewtau} constrain this parameter to $N_{\rm eff}=2.91_{-0.37}^{+0.39}$ at $95 \%$ c.l. in agreement with the standard model expectation of $N_{\rm eff}=3.046$ (corresponding to three active neutrinos) and with a $\sigma (N_{\rm eff}) \sim 0.2$ accuracy. 

While currently there is  no observational indication for a non-standard value of $N_{\rm eff}$ there are several physical mechanisms that can change its value. Sterile neutrinos \cite{sterile}, gravitational waves \cite{gw}, axions \cite{axions}, gravitino decays \cite{gravitinos}, and self-interacting dark matter \cite{selfneff} (just to name a few) can indeed all modify $N_{\rm eff}$. 

As discussed in \cite{baumaneff}, any particle that decouples from the primordial thermal plasma before the QCD transition will contribute with $\Delta N_{\rm eff}=N_{\rm eff}-3.046\ge 0.3$. This number has already been tested with Planck and near future data can fully falsify this hypothesis. If, however, the coupling happens after the QCD transition then any relativistic particle will contribute with a minimum value of $\Delta N_{\rm eff} =0.027$. More precisely the minimum contribution for a single real scalar particle is $\Delta N_{\rm eff} =0.027$, for a Weyl fermion it is $\Delta N_{\rm eff} =0.047$, and $\Delta N_{\rm eff} =0.054$ for a light vector boson (\cite{baumaneff}).

It is therefore clear that future CMB experiments reaching an experimental sensitivity of $\sigma (N_{\rm eff})=0.013$ will have the potential to rule out the existence of any relativistic particle beyond those predicted in the standard model at more than $95 \%$ c.l.. 

Reaching an accuracy on $N_{\rm eff}$ that is almost $15$ times better than current uncertainties is clearly an ambitious and difficult task. In \cite{S4} (see also \cite{core}) it has been shown that, in principle, CMB observations could reach this sensitivity, provided a perfect foreground removal, an angular resolution of $\sim 1'$, a sampled sky fraction above $f_{\rm sky}>0.6$, and a noise detector in temperature of $0.2$ $\mu K$-arcmin (see Figure 25 in \cite{S4}).

While the signal could indeed be present in the CMB sky, in this paper we highlight two key assumptions made in the forecasts that could undermine the possibility of reaching this sensitivity even for an ideal experiment.

The first assumption concerns inflation. In the forecasts made in \cite{S4} the power spectrum of primordial perturbations has been assumed to follow the usual power law form $P(k)=A_S k^{n_S}$ where $k$ is the perturbation comoving scale, $A_S$ and $n_S$ are the inflationary scalar amplitude and spectral index respectively (see e.g.\cite{Baumann:2009ds}).
However inflation generally predicts the presence of a running the spectral index $n_{\rm run}={\rm d}n_S/{\rm d}\ln k$ of the order of $(1-n_S)^2\sim 0.001$. Varying $n_{\rm run}$ produces similar effect in the CMB spectrum of a variation in the neutrino number. The inclusion of the running in the analysis introduces a degeneracy between $n_{\rm run}$ and $N_{\rm eff}$ that significantly weakens the achievable $\Delta N_{\rm eff}$. Moreover, an unaccounted negative running could mimic a value for $\Delta N_{\rm eff} >$ 0 suggesting the presence of new light particles.

The second assumption we investigate is related to the value of the primordial Helium abundance. It is well known that the Helium abundance parameter $Y_p$ is strongly degenerate with $N_{\rm eff}$ (see e.g. \cite{neff}). Letting $Y_p$ also vary freely in the analysis would greatly weaken the bounds on $N_{\rm eff}$. The most stringent forecasts presented in \cite{S4} or \cite{core} assume an Helium abundance derived from standard Big Bang Nucleosynthesis (BBN). However, even the most accurate BBN code, given a value of the baryon density and $N_{\rm eff}$, will produce an aestimate of $Y_p$ that is affected by a small uncertainty. This uncertainty mainly comes from the current experimental error on the neutron life time $\tau_n$. According to the latest Particle Data Group edition \cite{PDG}, the neutron lifetime is known with a precision of $\tau_n=880.2\pm 1.0$ s at $68 \%$ C.L. but this is an averaged value over different experimental constraints that are discrepant at the level of $4$ standard deviations. A larger, more conservative, uncertainty on the value of $\tau_n$ can therefore affect the precision in $Y_p$ and the final accuracy in $\Delta N_{\rm eff}$.

The goal of this paper is therefore to assess the impact of these two assumptions in future determinations of $N_{\rm eff}$ from CMB anisotropies. 

\section{Method}

\begin{table}
\begin{center}
\begin{tabular}{|c|c|c|c|c|c|}
\hline
Configuration                    & Beam &Power noise $w^{-1}$& $\ell_{max}$&$\ell_{min}$& $f_{\rm sky}$\\
\hline
Stage-IV      & $3$' & $1$ ($\mu K$-arcmin)& $3000$&$5$& $0.4$\\
\hline
Stage-IV+      & $1$' & $0.5$ ($\mu K$-arcmin) & $5000$&$2$& $0.4$\\
\hline
\end{tabular}
\end{center}
\caption{Experimental specifications for the two configurations considered in the forecasts.}
\label{tab:specifications}
\end{table}

In this paper, we forecast the ability of future CMB experiments to constrain the effective neutrino number $N_{\rm eff}$ in different theoretical frameworks.

Following the now common approach already used, for example, in \cite{core}, we perform Monte Carlo Markhov Chains (MCMC) analyses on mock data for several possible future experimental configurations, assuming a fiducial, vanilla, flat $\Lambda$CDM  model compatible with the recent Planck 2015 results \cite{plancknewtau}. More specifically, we assume a baryon and cold dark matter densities of $\Omega_{b}h^2=0.02225$ and $\Omega_{c}h^2= 0.1198$, an optical depth $\tau=0.0596$, an inflationary spectral index $n_s=0.9645$, and $3$ neutrinos with effective number $N_{\rm eff}=3.046$. While we consider $\Lambda$CDM as fiducial model, in our MCMC analysis we let vary also the neutrino effective number $N_{\rm eff}$ and the running $n_{\rm run}$. 

We use the publicly available Boltzmann code, CAMB~\cite{camb} to compute the theoretical CMB angular power spectra $C_{\ell}^{TT}$, $C_{\ell}^{TE}$, $C_{\ell}^{EE}$, $C_{\ell}^{BB}$ for temperature, cross temperature-polarization and $E$ and $B$ modes polarization \footnote{We neglect the non-gaussianity of the lensed $B$ modes and we do not delense. Our assumptions are therefore slightly more conservative than those presented in \cite{S4} where also simulated Planck data was considered.}.

In our simulations we make use of an instrumental noise given by the usual expression:

\begin{equation}
N_\ell = w^{-1}\exp(\ell(\ell+1)\theta^2/8\ln2),
\end{equation}

\noindent where $\theta$ is the experimental FWHM angular resolution, $w^{-1}$ is the experimental power noise expressed in $\mu K$-arcmin.
The total variance of the multipoles $a_{lm}$ is therefore given by the sum of the fiducial $C_\ell$'s with the instrumental noise $N_\ell$.

We consider two experimental configurations: a Stage-IV experiment as in \cite{calabresestage4} and a futuristic/optimistic "Stage-IV+" configuration with improved angular resolution and sensitivity as suggested in \cite{S4}.  We generate fiducial and noise spectra with noise properties as reported in Table \ref{tab:specifications}. Since we are mainly interested here in the impact of theoretical assumptions we assume negligible beam uncertainties and no foreground contaminations. However, we limit the temperature and polarization power spectrum from Stage-IV data in the range $5\le \ell \le3000$. As we report in the next section with the Stage-IV configuration reported in Table \ref{tab:specifications} we get forecasts on $N_{\rm eff}$ with uncertainties that are about $\sim 20 \%$ larger than those reported in \cite{S4}. For the more optimistic Stage-IV+ configuration we consider $\ell_{\rm max}=5000$. For both configurations we consider a sampled sky fraction of $f_{\rm sky}=0.4$. We do not include simulated Planck data with $f_{\rm sky}=0.2$ as in \cite{S4}. 

The simulated experimental data are then compared with a MCMC method with a theoretical model assuming a Gaussian likelihood ${\cal L}$ given by 
\begin{equation}
 - 2 \ln {\cal L} = \sum_{l} (2l+1) f_{\rm sky} \left(
\frac{D}{|\bar{C}|} + \ln{\frac{|\bar{C}|}{|\hat{C}|}} - 3 \right),
\label{chieff}
\end{equation}
where $\bar{C}_l$ and $\hat{C}_l$ are the assumed fiducial
and theoretical spectra plus noise and 
$|\bar{C}|$, $|\hat{C}|$ are the determinants of
the theoretical and observed data covariance matrices given by:
\begin{eqnarray}
|\bar{C}| &=& \bar{C}_\ell^{TT}\bar{C}_\ell^{EE}\bar{C}_\ell^{BB} -
\left(\bar{C}_\ell^{TE}\right)^2\bar{C}_\ell^{BB} ~, \\
|\hat{C}| &=& \hat{C}_\ell^{TT}\hat{C}_\ell^{EE}\hat{C}_\ell^{BB} -
\left(\hat{C}_\ell^{TE}\right)^2\hat{C}_\ell^{BB}~,
\end{eqnarray}
where $D$ is
\begin{eqnarray}
D  &=&
\hat{C}_\ell^{TT}\bar{C}_\ell^{EE}\bar{C}_\ell^{BB} +
\bar{C}_\ell^{TT}\hat{C}_\ell^{EE}\bar{C}_\ell^{BB} +
\bar{C}_\ell^{TT}\bar{C}_\ell^{EE}\hat{C}_\ell^{BB} \nonumber\\
&&- \bar{C}_\ell^{TE}\left(\bar{C}_\ell^{TE}\hat{C}_\ell^{BB} +
2\hat{C}_\ell^{TE}\bar{C}_\ell^{BB} \right). \nonumber\\
\end{eqnarray}
For our MCMC runs we use the publicly available Markov Chain Monte Carlo package {\sc CosmoMC}\footnote{\tt http://cosmologist.info}~\cite{Lewis:2002ah} sampling parameters with the Metropolis-Hastings algorithm, with a convergence diagnostic based on the Gelman and Rubin statistic.

\section{Results}
In this section we present our findings, discussing the impact of the two mentioned assumptions in the determination of $N_{\rm eff}$.

\subsection{Impact of the running of the spectral index}
We first analyze the case of a possible presence of running of the inflationary spectral index. 
We remind that the slow-roll solution for the primordial power spectrum can be expressed as (see e.g. \cite{cabass}):
\bea
  1 - \ns &=& 2\epsilon-\frac{\epsilon_{,N}}{\epsilon}-\frac{c_{\mathrm{s},N}}{c_\mathrm{s}} \label{eq:eom1} \\
     &=& \frac{r}{8 c_\mathrm{s}} - \frac{r_{,N}}{r}\,\,, \label{eq:eom2} \\
     n_{\rm run} &=& 2\epsilon_{,N}-\frac{r_{,NN}}{r} + \bigg(\frac{r_{,N}}{r}\bigg)^{2}\,\,, \label{eq:eom3}
\eea
where $\epsilon$ is the slow roll parameter, $c_\mathrm{s}$  is the inflaton speed of sound, $_{,N}$ refers to a derivative with respect to the number of e-foldings (see e.g. \cite{Baumann:2009ds}) and the tensor-to-scalar ratio is given approximately by $r=16\epsilon c_\mathrm{s}$.

Combining the above equations it is possible to write:
\begin{equation}
n_{\rm run} = (1-\ns)^{2} - 6\epsilon (1-\ns) + 8\epsilon^{2} - \bigg(\frac{rc_\mathrm{s,N}}{8c_\mathrm{s}^2}+\frac{r_{,NN}}{r}\bigg)\,\,.
\end{equation}
In a typical slow-roll model $n_{\rm run}$ is therefore {\it naturally} expected to be of the same order of $(1-\ns)^{2}$. Assuming a value of $n_s=0.955$, compatible in between $2$ standard deviations with current constraints from Planck \cite{planck2015}, we have $(1-\ns)^{2}\sim 0.002$, that is approximately the same level of expected sensitivity on $n_{\rm run}$ for the Stage-IV experiment (\cite{S4}).

More specifically, if we consider the Starobinsky model \cite{Starobinsky:1980te} with $c_\mathrm{s} = 1$ and $\epsilon = 3/(4N^{2})$, we obtain 
\bea
  \left( 1-\ns \right)^{2} &\simeq& \frac{4}{N^{2}}\,\,, 
  \\
  \frac{r_{,NN}}{r} &=& \frac{6}{N^{2}}\,\,. 
  \eea
that corresponds to a value for the running (again for $n_s=0.955$) of:
\be
\label{eq:starob3}
\nrun\simeq -\frac{2}{N^{2}}\simeq -\frac{1}{2}(1-\ns)^{2}\,\,\simeq -0.001.
\ee
It will not possible for Stage-IV alone to detect the running in case of the Starobinsky model and several other models predict a similar running (e.g. see \cite{bellido}). However, as we discuss below, if not considered in the analysis it may anyway affect the constraints on other, correlated, parameters as $N_{\rm eff}$. Moreover, a larger running is expected in several theoretical scenarios as (just to name a few) a breakdown of the slow roll approximation \cite{peiris}, multiple fields inflation \cite{multifield}, the presence of a non-canonical kinetic term \cite{chung} and the running-mass models \cite{Covi:2004tp}.

\


\begin{table}
\begin{center}
\begin{tabular}{|c|c|c|}
\hline
Case                    & $N_{\rm eff}$ (Stage-IV) &$N_{\rm eff}$ (Stage-IV+) \\
\hline
Varying $n_{\rm run}$         & $3.049 \pm 0.076$& $3.048^{+0.023}_{-0.026}$\\
\hline
$n_{\rm run}=0$       & $3.048 \pm 0.043$ & $3.047 \pm 0.021$\\
\hline
$n_{\rm run}=0.002$       & $3.019 \pm 0.043$ & $3.035 \pm 0.021$\\
$n_{\rm run}=0.004$       & $2.996 \pm 0.044$ & $3.024 \pm 0.021$\\
$n_{\rm run}=-0.002$       & $3.074 \pm 0.044$ & $3.056 \pm 0.021$\\
$n_{\rm run}=-0.004$       & $3.098 \pm 0.044$ & $3.071 \pm 0.019$\\
\hline
\end{tabular}
\end{center}
\caption{Constraints at $68 \%$ c.l. for $N_{\rm eff}$ assuming different values for the running. Including the running in the analysis (first row) increases the error on $N_{\rm eff}$ by $\sim 76 \%$ for the Stage IV experiment ($\sim 17 \%$ for a Stage-IV+ experiment) respect to the no-running case (second row). Neglecting the running shifts the mean value by approximately $\Delta N_{\rm eff}\sim -12n_{\rm run}$ for Stage-IV and $\Delta N_{\rm eff}\sim -5n_{\rm run}$ for a Stage-IV+ experiment.}
\label{tablenrun}
\end{table}

\begin{figure}[!hbt]
\includegraphics[width=.48\textwidth]{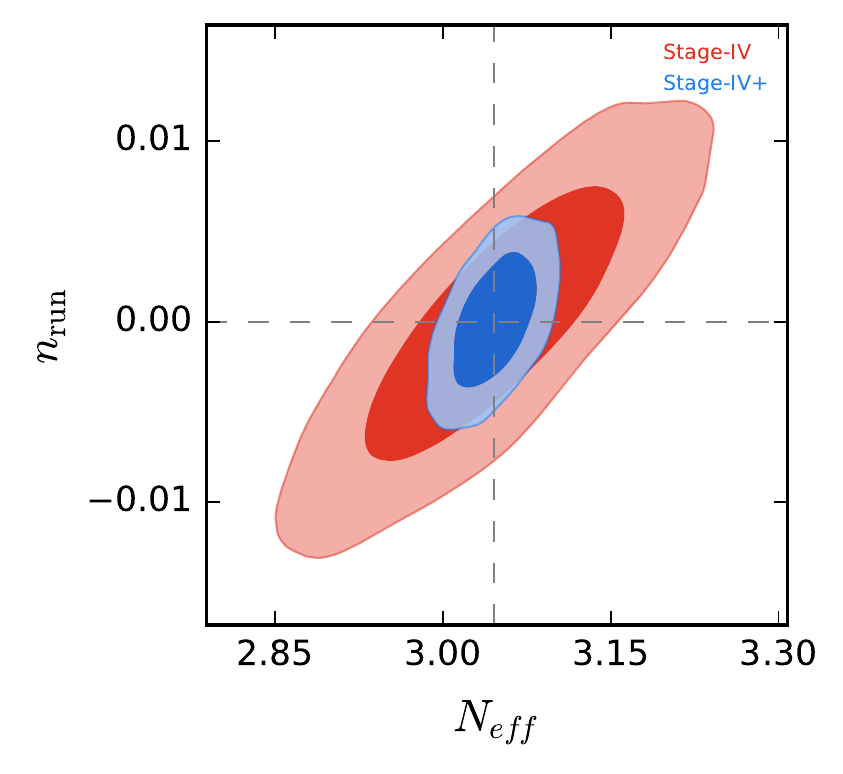}
\caption{Contour plots at the $68 \%$ and $95 \%$ confidence level forecasted on the $n_{\rm run}$ vs $N_{\rm eff}$ plane for the Stage-IV experiment and for a optimistic Stage-IV+ upgrade. A degeneracy between the two parameters is clearly present, more pronounced in the case of Stage-IV. }
\label{figure1}
\end{figure}

\begin{figure}[!hbt]
\includegraphics[width=.48\textwidth]{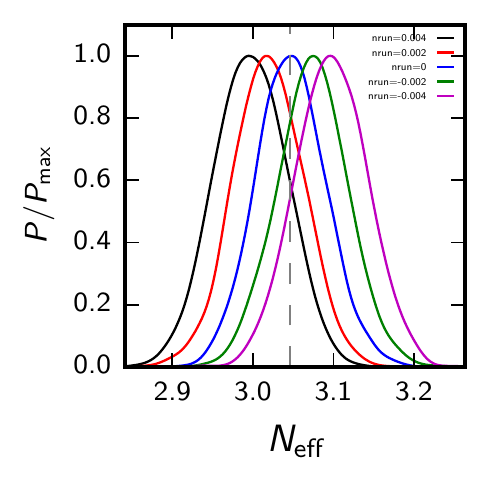}
\includegraphics[width=.48\textwidth]{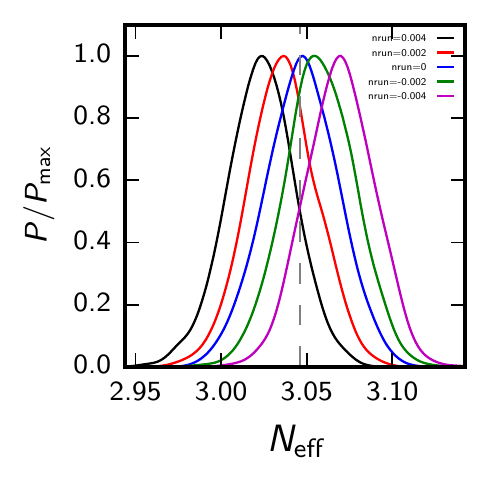}
\caption{Posterior distributions on $N_{\rm eff}$ assuming $N_{\rm eff}=3.046$ and different values for $n_{\rm run}$ for the fiducial model but performing an analysis with $n_{\rm run}=0$. Top Figure are the posteriors from Stage-IV while in the bottom we have the posteriors for a Stage-IV+ experiment. As we can see, not accounting for a negative running could produce a significant shift in the recovered values of $N_{\rm eff}$.}
\label{figure2}
\end{figure}

It is therefore interesting to perform an analysis on future mock data allowing both $N_{\rm eff}$ and $n_{\rm run}$ to vary.
The results of this analysis are reported in Table~\ref{tablenrun} and in Figure~\ref{figure1} and ~\ref{figure2}. As we can see from Figure~\ref{figure1} and from the first row of Table~\ref{tablenrun}, a strong degeneracy between the running of spectral index and the neutrino effective number exists. Namely, a decrease/increase of $N_{\rm eff}$ can be counterbalanced with a decrease/increase of $n_{\rm run}$. The main effect of this degeneracy is a significant increase in the forecasted uncertainty on $N_{\rm eff}$ when the $n_{\rm run}$ parameter is considered. Indeed, by comparing the constraints between the first two rows of Table~\ref{tablenrun} we see that the inclusion of $n_{\rm run}$ could result in a $\sim 76 \%$ decrease in the accuracy on $N_{\rm eff}$ for a Stage-IV experimental configuration. In practice opening the natural possibility of a running prevents the Stage-IV experiment to reach the goal of a $\sigma (N_{\rm eff}) \sim 0.03$ accuracy. The $n_{\rm run}$-$N_{\rm eff}$ degeneracy is less present in the case of a Stage-IV+ experiment. Still, when running is considered, the constraints on $N_{\rm eff}$ are weaker also in this case.

It is interesting to investigate what effect on $N_{\rm eff}$ could be produced by unaccounted running, i.e., when performing a MCMC analysis fixing $n_{\rm run}=0$ but adding non-zero running to the fiducial model. As we can see from Table~\ref{tablenrun} and the posteriors in Figure~\ref{figure2}, we found that unaccounted running could shift the mean value of $N_{\rm eff}$ from the standard value $N_{\rm eff}=3.046$ by
\begin{equation}
\Delta N_{\rm eff} \sim -12n_{\rm run}
\end{equation}
for a Stage-IV configuration and
\begin{equation}
\Delta N_{\rm eff} \sim -5n_{\rm run}
\end{equation}
for a Stage-IV+ experiment. As we can see, therefore, even a small {\it negative} running of $n_{\rm run}\sim -0.001$ could produce a {\it positive} shift of $\Delta N_{\rm eff} \sim 0.01$ in the recovered mean value of $N_{\rm eff}$ for Stage-IV. Moreover, one should consider that Stage-IV will have an accuracy of $\sigma(n_{\rm run})\sim0.002$ at $68 \%$ c.l.. Even with a value of $n_{\rm run}=0.001$, a statistical fluctuation of $\sim 1.5$ standard deviation could be possible,  yielding $n_{\rm run} \sim 0.004$ and a shift of $\Delta N_{\rm eff} \sim 0.048$. In practice, it will be hard for Stage-IV to discriminate between a negative running and the presence of a light vector boson.

Fortunately, the situation appears brighter when considering a  Stage-IV+ experiment. In this case, as we can see from Figure~\ref{figure1} and Figure ~\ref{figure2}, bottom panel, the degeneracy between $N_{\rm eff}$ and $n_{\rm run}$ is less significant. Without running, we found that a Stage-IV+ experiment could reach a sensitivity of $\Delta N_{\rm eff} =0.021$ (see Table~\ref{tablenrun}, excluding at more than $95 \%$ c.l. the minimum contribution of $\Delta N{\rm eff}=0.054$ for a light vector boson. However, in this case a {\it positive} running of $n_{\rm run}\sim 0.001$ could produce a {\it negative} shift of $\Delta N_{\rm eff} \sim -0.005$. This would be enough for bringing such signal back in agreement between the $2$ standard deviation threshold. 

\subsection{Impact of the neutron lifetime}

The second key assumption we want to investigate concerns the value of the neutron lifetime. The forecasts on $N_{\rm eff}$ presented in \cite{S4} and in the previous section generally assume a value on the primordial Helium abundance derived from Big Bang Nucleosynthesis. In practice, for each theoretical model, a value for $Y_p$ is obtained through a BBN code ("Parthenope", \cite{naples}) assuming the baryon density $\omega_b$ of the model and a neutron lifetime of $\tau_n=880.3$ s. An accurate determination of the Helium abundance $Y_p$ is crucial for the determination of $N_{\rm eff}$.
The two parameters are indeed correlated and without the assumption of BBN the accuracy on $N_{\rm eff}$ is larger than $\sigma (N_{\rm eff}) = 0.07$ even for the most optimistic experimental configuration (see \cite{S4}).
However the neutron lifetime is known with an experimental error. As it is well known, this uncertainty propagates in a systematic error on the BBN derived value of $Y_p$ that may affect the constraints on $N_{\rm eff}$.
Given a value of the neutron lifetime, one could expect from a numerical fit a helium abundance given by:
\begin{equation}
Y_p(\tau_n)=({{\tau_n}\over{880.3s}})^{0.73}Y_p(\tau_n=880.3s),
\label{tauyp}
\end{equation}
where $Y_p(\tau_n=880.3s)$ is the Helium abundance derived assuming $\tau_n=880.3$ s.

The most recent bound on the neutron lifetime from the Particle Data Group (PDG) is \cite{PDG}:
\begin{equation}
\tau_n=(880.2\pm1.0) s
\end{equation}
Assuming therefore a $2$ standard deviations fluctuation we could expect from Eq.~\ref{tauyp} an increase in $Y_p$ of $\sim 0.18 \%$. This is about $\sim 15 \%$ of the expected Stage-IV accuracy on $Y_p$ of  $\sigma(Y_p)=0.025$ (for $Y_p=0.2466$) and it does not therefore significantly affect the Stage-IV future constraint on $N_{\rm eff}$.

However there is a longstanding tension between different measurements of the neutron lifetime based on different experimental techniques (see e.g. \cite{tauntension,tauntension2,tauntension3}).
The most recent measurements of the neutron lifetime are indeed
based on two different experimental techniques: counting 
the $\beta$-decay products in a passing beam of cold neutrons ("beam" method) or counting the ultra cold neutrons (UCN) surviving in a storage bottle ("bottle" method).

The most recent value from the "beam" method is \cite{yue}:
\begin{equation}
\tau_n^{beam}=(887.7 \pm 1.2 [Stat] \pm 1.9 [Sys]) s,
\end{equation}
while the most precise measurement using the "bottle" method is
\cite{serebrov}:
\begin{equation}
\tau_n^{bottle}=(878.5 \pm 0.7 [Stat] \pm 0.3 [Sys]) s.
\end{equation}
These two measurements (summing the systematic errors in quadrature) are therefore discrepant at the level of $\sim 3.9$ standard deviations. Moreover, the beam determination is in tension at the level of $\sim 3$ standard deviations with the quoted PDG constrain, while also the "bottle" constraint shows some tension albeit just at  $1.4$ standard deviations.

Recently, a new measurements with an improved "bottle" method has been reported \cite{pattye}, giving :
\begin{equation}
\tau_n^{bottle}=(877.7 \pm 0.7 [Stat] \pm 0.3/-0.1 [Sys]) s.
\end{equation}
in tension with the PDG value at $\sim 2$ standard deviations. 

Given the large inconsistencies between experimental values, it makes certainly sense to investigate how a larger uncertainty on $\tau_n$ than the one quoted in the PDG could impact future CMB constraints on $N_{\rm eff}$. 

We therefore simulate future CMB data assuming standard $\Lambda$CDM but with two possible "real" values for the neutron lifetime: $\tau_n^{high}=888.0$ s, consistent with "beam" measurements, and $\tau_n^{low}=877.0$ s, consistent with "bottle" experiments, corresponding to different values for the BBN derived primordial Helium abundance. We then analyze these datasets assuming the quoted PDG value of $\tau_n=880.2$s, recovering the value of $N_{\rm eff}$ and quantifying the possible bias introduced by a wrong assumption on $\tau_n$ (and on the primordial Helium abundance $Y_p$).

\begin{table}
\begin{center}
\begin{tabular}{|c|c|c|}
\hline
Case                    & $N_{\rm eff}$ (Stage-IV) &$N_{\rm eff}$ (Stage-IV+) \\
\hline
$\tau_n=880.2$s       & $3.048 \pm 0.043$ & $3.047 \pm 0.021$\\
\hline
$\tau_n=888.0$s       & $3.062 \pm 0.040$ & $3.064 \pm 0.021$\\
$\tau_n=877.0$s       & $3.039 \pm 0.041$& $3.037 \pm 0.020$\\
\hline
\end{tabular}
\end{center}
\caption{Constraints at $68 \%$ C.L. for $N_{\rm eff}$ assuming different values for the neutron lifetime.}
\label{tabletaun}
\end{table}

The results are reported in Table \ref{tabletaun}. As we can see,
an unaccounted higher value for the neutron lifetime $\tau_n^{high}$ could introduce a bias in the neutrino effective number of $\Delta N_{\rm eff} \sim 0.015$ while a lower value $\tau_n^{low}$ could bring to a $\Delta N_{\rm eff} \sim -0.009$. In both the Stage-IV and Stage-IV+ cases this possible systematic will not affect the $N_{\rm eff}$ constraint in a significant way such to mimic a detection at more than two standard deviation. However, it may make a statistical fluctuation more significant than what it actually is and, conversely, reduce the significance of a real discovery. In concreto, assuming a conservative experimental uncertainty on
the neutron lifetime of $\sim 10$s introduces a systematic error of $|\Delta N_{\rm eff}|\sim 0.02$, placing a serious limitation to the ultimate goal of $\sigma (N_{\rm eff})=0.013$.

\section{Conclusions}

In this paper, we have considered the impact of two theoretical assumptions in the forecasted accuracy on the neutrino effective number for future CMB experiments. As illustration, we have considered two experiments: the future Stage-IV experiment and a further, optimistic, upgrade to a Stage-IV+ experiment. The first assumption concerns the running of the inflationary spectral index, usually assumed as negligible. The second assumption is related to a perfect knowledge of the neutron lifetime. Both assumptions are not particularly well motivated: slow roll inflation predicts a running of the same order of the accuracy expected from future experiments, while a $\sim 4 \sigma$ tension between current experimental measurements of $\tau_n$ is present, potentially suggesting a significantly larger systematic error. 

We found that for the Stage-IV experiment a running of $n_{\rm run}\sim -0.002$ $(n_{\rm run}\sim -0.004$) could result in a {\it positive} shift of $\Delta N_{\rm eff} \sim 0.03$ ($\Delta N_{\rm eff} \sim 0.05$). Running could therefore strongly impact the abilities of the Stage-IV experiment to significantly rule out or detect the presence of an extra relativistic particle at recombination. In case of an highly optimistic Stage-IV+ experiment, the correlation between running and $N_{\rm eff}$ is less significant and the results are less affected.

When considering the neutron lifetime we found that if we assume the current uncertainties reported in the PDG then the impact is minimal. However, in the case of a different, larger, value for $\tau_n$, compatible with current "beam" measurements, or smaller, compatible with the most recent "bottle" experiments, we found a shift of $\Delta N_{\rm eff} \sim 0.016$ and $\Delta N_{\rm eff} \sim -0.008$ for both Stage-IV and Stage-IV+. While we are clearly considering a very conservative uncertainty on $\tau_n$, nearly ten times larger than what reported in the PDG, we have also to bear in mind that any claim of new physics from the CMB must withstand a severe scrutiny of the assumptions made.

Both running and current experimental uncertainties on $\tau_n$ can therefore undermine the possibility of reaching the accuracy of $\Delta N_{\rm eff}\sim 0.013$ needed for ruling out the presence of any extra relativistic particle at more than $95 \%$ c.l.. Moreover, current uncertainties on the value of the neutron lifetime also limit the accuracy on $n_{run}$ achievable from future CMB experiments.

However, complementary cosmological observables as galaxy surveys (see e.g. \cite{surveys}), 21 cm line fluctuations \cite{21cm,21cmb} and, possibly, CMB spectral distortions (see e.g., \cite{Chluba2012_2x2, cabass}) could help in breaking the degeneracy between $n_{\rm run}$ and $N_{\rm eff}$. At the same time, new experiments expected in the next years will be crucial in solving the current neutron lifetime discrepancy \cite{taunfuture}.

Before concluding we want to point that in this paper we discussed just two possible assumptions that can bias the derived value for $N_{\rm eff}$. Other extensions of the standard model can produce similar effects. We plan to further analyze these extensions in a future paper \cite{cappa2}. We also confirmed that the remaining theoretical uncertainties between different standard recombination codes (e.g., {\tt CosmoRec} \citep{Chluba2010} and {\tt Recfast} \citep{Seager2000, Wong2008}) produce effects that are below what was found here, even when including refined helium recombination physics \citep{Chluba2012}.

\

\acknowledgments 
EDV acknowledges support from the European Research Council in the form of a Consolidator Grant with number 681431. AM thanks the University of Manchester and the Jodrell Bank Center for Astrophysics for hospitality. JC is supported by the Royal Society as a Royal Society University Research Fellow at the University of Manchester and the European Research Council through Consolidator Grant, No. 725456, UK.

\end{document}